\documentclass[12pt]{iopart}

\newcommand{\Sh}{Schr\"odinger{ }}

\newcommand{\Fig}[1]{Fig. \ref{#1}}
\newcommand{\braket}[3]{\left\langle #1 \left\vert #2
            \right\vert #3 \right\rangle}
\newcommand{\brakete}[3]{\left\langle #1 \right\vert #2
            \left\vert #3 \right\rangle}
\newcommand{\brak}[2]{\left\langle #1 \left\vert
             #2 \right. \right\rangle}

\newcommand{\fues}[1]{\left(#1\right)}

\newcommand{\yav}[1]{\left[#1\right]}

\newcommand{\llal}[1]{\left\{#1\right.}
\newcommand{\abs}[1]{\left\vert#1\right\vert}

\newcommand{\Eq}[1]{Eq. (\ref{#1})}
\newcommand{\Eqs}[1]{Eqs. (\ref{#1})}

\usepackage{epsfig}  %% To include eps figures 
\usepackage{comment} %% For large regions of commented text
\usepackage{amssymb} %% For the AMS symbols

\def\lambdabar{\protect\@lambdabar}
\def\@lambdabar{%
\relax
\bgroup
\def\@tempa{\hbox{\raise.73\ht0
\hbox to0pt{\kern.25\wd0\vrule width.5\wd0
height.1pt depth.1pt\hss}\box0}}%
\mathchoice{\setbox0\hbox{$\displaystyle\lambda$}\@tempa}%
{\setbox0\hbox{$\textstyle\lambda$}\@tempa}%
{\setbox0\hbox{$\scriptstyle\lambda$}\@tempa}%
{\setbox0\hbox{$\scriptscriptstyle\lambda$}\@tempa}%
\egroup
}

\begin{document}  

\title[Photoconductivity in  Ac-driven  lateral superlattice in the presence of a magnetic field.]
      {Photoconductivity in  Ac-driven  lateral superlattice in the presence of a magnetic field.}    

\author{Manuel Torres}
\address{ Instituto de F\'{\i}sica,
Universidad Nacional Aut\'onoma de M\'exico, Apartado Postal
20-364,  M\'exico Distrito Federal 01000,  M\'exico}
\ead{torres@fisica.unam.mx}

\author{Alejandro Kunold}
\address{Departamento de Ciencias B\'asicas, Universidad Aut\'onoma
Metropolitana Azcapotzalco, Av. San Pablo 180,  M\'exico Distrito Federal  02200, M\'exico}
\ead{akb@correo.azc.uam.mx}

\date{\today}

%%% Abstract %%%   
\begin{abstract}  
    
In this work we present a  model for the photoconductivity of two-dimensional electron system in a perpendicular homogeneous magnetic field,  a weak lateral   superlattice, and  exposed to 
millimeter irradiation.  The model includes the microwave and Landau contributions in a non-perturbative  exact way, the periodic potential is  treated perturbatively.  The Landau-Floquet states   provide a convenient base with respect to which the lattice  potential becomes  time-dependent,  inducing   transitions between the Landau-Floquet  levels.  Based on this formalism, we provide a  Kubo-like   formula that takes into account the oscillatory Floquet structure of the problem.    The  total conductivity exhibits strong oscillations, determined by
  $\epsilon = \omega /  \omega_c$ with $\omega$ the radiation frequency 
and $\omega_c$ the cyclotron frequency. The oscillations follow a pattern with minima
centered  at $\omega/\omega_c =j + \frac{1}{2} (l-1) +  \delta $,  and maxima centered at  $\omega/\omega_c =j + \frac{1}{2} (l-1) -  \delta $, where $j=1,2,3.......$,   $\delta$ is a constant phase shift and $l$ is the dominant multipole  contribution. 
Negative conductance states develop as the electron mobility and the  intensity of the microwave power  are increased. 
It is proposed that,  depending on the geometry,  negative conductance  sates or negative resistance states may be observed  in lateral superlattices  fabricated  in  $GaAs/AlGa As$  heterostructures.

\end{abstract}


%%%%%%%%%%%%%%%%%%%%%
%        Section 
%%%%%%%%%%%%%%%%%%%%%%%%
\section{Introduction.}\label{intro}
%%%%%%%%%%%%%%%%%%%%%

 Photon-assisted tunneling is a well established phenomenon, it is originated from the quasiparticle tunneling 
 in semiconductors irradiated by high frequency fields \cite{tien,bastard}. It has also been observed in high frequency transport in  semiconductor multiquantum well superlattices \cite{guimaraes,keay1,keay2} and nanostructures 
 \cite{verg,kou}.  Photon assisted tunneling also  plays an important role in the development of an intersubband 
 laser \cite{faist}.  More recently Keay 
  $et \, al$ \cite{keay3,keay4}  reported 
 the observation of absolute negative conductance and multiphoton  stimulated emission in sequential resonant tunneling semiconductor   superlattices subjected to intense terahertz electric fields. Theoretical studies predicting dynamical localization and absolute negative conductance in semiconductor superlattices subjected to ac electric fields have been  known  for some time. These studies  are based in semiclassical models of electron motion in superlattices in the miniband or coherent tunneling regime \cite{platero}. 
 
On the other hand, the irradiation  with lower frequency fields of two-dimensional  electron systems $(2DES)$  has remarkable 
consequences on  the transport properties at low magnetic fields.   Recently, two experimental groups
\cite{zudov1,mani1,zudov2,mani2},  reported that high mobility $GaAs/Al_xGa_{1-x} As$  heterostructures  exposed to radiation from $20$ to $150 \, GHz$  resulted in 
giant magnetoresistance oscillations, periodic in $\epsilon = \omega /  \omega_c$ with $\omega$ the radiation frequency 
and $\omega_c$ the cyclotron frequency.  In the case of high mobility samples the minima of the magnetoresistance oscillations  
can form zero resistance  sates $(ZRS)$, the series of minima formed at $ \epsilon = j + \delta$, $j=1,2,3....$, $\delta = \frac{1}{2}$  \cite{zudov1,zudov2}, or  $\delta = \frac{1}{4}$ \cite{mani1,mani2}.
 Two distinct mechanism for photoconductivity  corrections have been proposed: $(i)$ the impurity scattering mechanism, which is caused by the 
 excitation of electrons by the combined effect  of photon and impurity scattering  
 \cite{ry1,ry2,durst,andre,shi,lei,vavilov,tor1}, and
$(ii)$ the distribution function mechanism, which involves redistribution of intra-Landau level population for large inelastic lifetimes 
\cite{doro,dmi1,dmi2,kennett}.  A model  for  the impurity scattering mechanism was proposed previously by the authors  \cite{tor1,tor2}, the model includes the microwave and Landau contributions in a non-perturbative  exact way, impurity scattering   effects are treated perturbatively, the model reproduce various of the experimentally observed features, in particular the fact that negative resistance states $(ZRS)$  appear only when the electron   mobility
exceeds  a threshold.

Experimental investigations  of lateral superlattices have found interesting  commensurability phenomena for the  magnetoresistance,
both for  weak potential modulations  
\cite{weiss-vk}   and also for  strong potential modulations \cite{schuster}.  The observed phenomena  can be related  to the 
commensurability  of the ratio  of the classical cyclotron diameter to the lattice constant, or by particular chaotic trajectories.  Lateral superlattices fabricated by cleaving  a  square grid of cylindrical holes with periodicity  $\approx \,  100 \, to \, 120 \, nm$ 
 on a $GaAs/AlGaAs$ heterostructure have been developed by von Klitzing group \cite{klitbut} in order to provide 
 evidence  of the Hofstadter fractal energy spectrum in the quantized Hall conductance.
 
In the present  work we propose that 
 the   combined effect of a perpendicular magnetic field plus the  irradiation of lateral    superlattices can give rise to interesting oscillatory conductance phenomena, with the possible 
 development of negative conductance states $(NCS)$. The observation of NCS  would require a Corbino sample,   alternatively negative resistance states $(NRS)$ would require a   Hall  geometry.   We consider a
 2DES in the presence of a lateral  superlattice and subjected to  magnetic and microwave fields.  As a first step, it is shown that  the  dynamics associated with the Landau and radiation contributions   can be exactly taken into account.  As a second step, the periodic potential is added  perturbatively. With respect to the  Landau-Floquet states, the periodic potential acts as  a coherent oscillating field which  induces   transitions between these levels.   Based on this formalism,  we provide  a  Kubo-like  expression for the conductance that incorporates  the   oscillatory Floquet structure of the system and the   impurity scattering in the usual Born approximation.  
 It is found  that  $\sigma_{xx}$  exhibits strong oscillations determined by 
  $\epsilon = \omega /  \omega_c$. NCS develop for sufficiently high  electron mobility and  strong microwave power. 
   The model is  used to test chirality  effects induced by the magnetic field, calculations are carried out  for various $\mathbf{E}$-field polarization's.  Finally,  we explore the nonlinear regime in which multiple photon exchange  plays an essential role, as well as the current-voltage characteristics of the system. 
   
 The paper is organized as follows. In the next section we present
the model and the   method that allow us to obtain the exact solution of the Landau-microwave
system, as well as  the perturbative corrections induced by the periodic  potential. 
 In section \ref{kubo} we develop the formulation of the  linear response theory  valid in arbitrary  magnetic and microwave  fields.    A discussion of relevant numerical calculations is presented in section
 \ref{results}. The last section contains a summary of our main results.

\section{The Model.}\label{model}
Let us consider the motion of an  electron in two dimensions subject to a uniform magnetic  field 
 $\mathbf{ B}$  
perpendicular to the plane and a constant electric field $\mathbf{E }_c$, a periodic potential $V$   and driven by  microwave radiation. On the plane the 
 dynamics is governed by the \Sh equation 
%%%%%%%%%%%%%%%%%%%%%%%%%%%%%%%%%%%
\begin{equation}\label{ecs1}
i \hbar \frac{\partial \Psi }{\partial t}= H \Psi  =  \left[  H_{\{B,\omega\}}  +   V ( \mathbf{r} )  \right] \Psi  \, .
\end{equation}
Here   $H_{\{B,\omega\}}$ is written in terms of   the covariant derivative 
%%%%%%%%%%%%%%%%%%%%%
\begin{equation}\label{ham1}
H_{\{B,\omega\}} = \frac{1}{2m^*} \mathbf{\Pi}^2 \, , \hskip2.0cm   \mathbf{\Pi} =   \mathbf{p}+e 
  \mathbf{A} \, , 
\end{equation}
where $m^*$ is the effective electron mass  over the plane that takes into account the effects of the crystalline atomic structure over the charge carriers. The vector potential  $\mathbf{A}$  includes  all the contributions  of:   the   magnetic, electric   and     radiation fields: 
 
 %%%%%%%%%%%%%%%%%%%%%%%%%%
 \begin{equation}\label{vecpot}
 \mathbf{A}  = - \frac{1}{2} \mathbf{r} \times \mathbf{B}  +  Re \,\, \left[  \frac{ \mathbf{\epsilon} E_\omega} {\omega} \exp\{ -i \omega t \} \right] \, +  \mathbf{E}^c \, t \,  . 
\end{equation}
  % %%%%%
The  superlattice  potential   $V(\mathbf{r}) $   is  decomposed in a Fourier expansion
%%%%%%%%%%%%%%%%%%%%%%%%%%%%%%%%
\begin{equation}\label{potimp}
 V(\mathbf{r})= \sum_{m \, n}  V_{m \, n} \exp \bigg{\{i}  2 \pi  \left( \frac{m \, x}{a} + \frac{n \,  y}{b}   \right) \bigg { \}} \, .
\end{equation}

     We first consider the exact solution of the microwave driven Landau problem, the periodic potential  effects are  lately 
 added perturbatively. This approximation is justified  when the following holds:  $(i)$:   $ \vert  V \vert /  \hbar \omega_c << 1$ and   $(ii)$: $\omega \, \tau_{tr}   \sim  \omega_c \, \tau_{tr}  >>  1$; $\tau_{tr}$  is the transport  relaxation time that is estimated using its relation to the   electron mobility $\mu = e \tau_{tr} / m^*$. 
 
 The system  posed by $H_{\{B,\omega\}}$  can be recast as a forced  harmonic oscillator,  a 
 problem that was solved long time ago by Husimi \cite{husi}.  Following the formalism developed in references \cite{Kunold1,Torres1}, we introduce a canonical transformation to new variables 
$Q_\mu,P_\mu$; $\mu=0,1,2$,  according to 
\begin{eqnarray}\label{trancan}
Q_0  &=&  t   \, ,   \hskip5.6cm  P_0 = i \partial_t + e \phi + e  \mathbf{r} \cdot  \mathbf{E}  ,\nonumber \\
\sqrt{e B} Q_1  &=& \Pi_y   \,,  \hskip4.4cm  \sqrt{e B}P_1 =\Pi_x ,\nonumber \\
\sqrt{e B} Q_2  &=& p_x+ e A_x + e By  \, ,   \hskip2.0cm  \sqrt{e B} P_2=p_y+ e A_y - e B x.
\end{eqnarray}
It is easily verified that the transformation is indeed  canonical, 
the  new variables obey  the commutation rules:  $- \left[Q_0,P_0\right]=  \left[Q_1,P_1\right]= \left[Q_2,P_2\right]=i B$; all other commutators being zero. The inverse transformation gives
$x=  l_B  \left(Q_1-P_2 \right)$, and  $ y=  l_B   \left( Q_2-P_1 \right) $, 
where $l_B=  \sqrt{\frac{\hbar}{eB}}$ is      the magnetic length. 
 The operators $( Q_2 , P_2)$ can  be  identified with the generators of the electric-magnetic translation symmetries \cite{Ashby1,Kunold2}. Final results are independent  of the selected gauge. 
>From the operators in \Eq{trancan}   we construct two pairs of
harmonic oscillator-like ladder operators: $(a_1,a_1^{\dag})$, and $(a_2,a_2^{\dag})$ with:
\begin{equation}\label{opasc}
a_1 =   \frac{1}{\sqrt{2}}    \left(P_1-iQ_1\right), \hskip2.0cm    a_2 =  \frac{1}{\sqrt{2} }    \left(P_2 -i Q_2\right), 
\end{equation}
obeying:  $[a_1,a_1^{\dag}]=[a_2,a_2^{\dag}]=1$, and $[a_1,a_2]=[a_1,a_2^{\dag}]=0$.

It  is now  possible to find a unitary transformation that exactly diagonalizates    $H_{\{B,\omega\}}$, it yields
 %%%%%%%%%%%%%%%%%%%%%%%%%%
\begin{equation}\label{tran1}
W^{\dagger} H_{\{B,\omega\}}  W = \omega_c \left( \frac{1}{2} + a_1^\dagger \, a_1\right) \equiv H_0 \, , 
\end{equation}
with the cyclotron frequency $\omega_c = eB/m^*$ and the $W(t)$ operator     given by 
 %%%%%%%%%%%%%%%%%%%%%%%%%%
\begin{equation}\label{opw}
W(t)=  \exp\{i \eta_1 Q_1\}  \exp\{i \xi_1 P_1\} \exp\{i \eta_2 Q_2\}  \exp\{i \xi_2 P_2\}   \exp\{i  \int^t  {\mathcal L} dt^\prime \} 
\, , 
\end{equation}
where  the functions $\eta_i(t)$ and $\xi_i(t)$ represent the solutions to the classical equations of motion that follow from the variation of the Lagrangian 
\begin{equation}\label{lagrang}
{\mathcal L}  =  \frac{\omega_c}{2} \left( \eta_1^2 + \zeta_1^2 \right) +    \dot \zeta_1\eta_1 +  \dot \zeta_2 \eta_2 
+  e l_B \,    \left[ E_x  \left( \zeta_1 + \eta_2 \right)   +    E_y   \left(\eta_1 + \zeta_2  \right) \right]     \, .
\end{equation}
%%%%%%%%%%%%%
 It  is  straightforward  to obtain the solutions to the equation of motion, using  the expression for the electric  field 
 $ \mathbf{E} = -\partial \mathbf{A} / \partial t  $ with $ \mathbf{A}$ given   in  (\ref{vecpot}). Adding   a damping term that takes into account the radiative decay of the quasiparticle, they read 
  %%%%%%%%%%%%%%%%%%%%%%%
 \begin{eqnarray}\label{soleqm}
\hskip-2.0cm & & \eta_1 =  e l_B  E_\omega  \, Re \left[\frac{-i \omega \epsilon_x + \omega_c \epsilon_y  }{\omega^2 - \omega_c^2 + i \omega \Gamma_{rad} }
     e^{i \omega t} \right]  ,   \hskip0.8cm     \eta_2 = e l_B E_\omega \,  Re \left[\frac{\epsilon_y e^{i\omega t}}{i \omega} \right] + e l_B E^c_y  \, t  ,                  \nonumber \\
\hskip-2.0cm & & \zeta_1 = e l_B  E_\omega \, Re \left[\frac{ \omega_c \epsilon_x + i \omega \epsilon_y  }{\omega^2 - \omega_c^2 + i \omega \Gamma_{rad}  }
     e^{i \omega t} \right] ,  \hskip0.8cm      \zeta_2 = -e l_B E_\omega \, Re \left[\frac{\epsilon_x e^{i\omega t}}{i \omega} \right] 
     - e l_B E^c_x  \, t   .
\end{eqnarray}
%%%%%%%%%%%%%%%%%%%%%%%%%%%%%%%%
According to the Floquet theorem the wave function  can be written as
 $\Psi (t) = \exp \left(- i {\cal E} _\mu  t \right) \phi_\mu(t)$, where $ \phi_\mu(t)$ is periodic in time, 
 $i.e.$ $ \phi_\mu(t + \tau_\omega ) =  \phi_\mu(t)$.  From   \Eq{opw} it is noticed that  the transformed wave function  $\Psi^W = W \Psi$ contains the phase factor $\exp \left( i \int^t   {\mathcal L}  d t^\prime \right) $. It then follows  that the quasienergies and the  Floquet modes can be deduced if we add and subtract to this  exponential  a term of the form $\frac{t}{\tau} \int_0^\tau {\mathcal L} dt^\prime$. Hence, the quasienergies can be  readily read off
%%%%%%%%%%%%%%%%%%%%%%%%
\begin{equation}\label{enerflo}
\hskip-2.0cm  {\cal E} _\mu ={\cal E} _\mu^{(0)} +   {\cal E}_{rad} \, ;  \hskip0.6cm  
   {\cal E} _\mu^{(0)} = \hbar \omega_c \left( \frac{1}{2} + \mu  \right) ,   \hskip0.6cm  {\cal E}_{rad} = 
   \frac{e^2 E_\omega^2 \left[ 1 + 2 \omega_c {Re(\epsilon_x^* \epsilon_y) }/ \omega\right] } {2 m^* \left[ \left(\omega - \omega_c \right)^2 + \Gamma_{rad}^2 \right]}, 
\end{equation}
here  ${\cal E} _\mu^{(0)}$ are   the usual Landau energies, and the induced  Floquet energy shift is given by the microwave energy 
$ {\cal E}_{rad} $.   
The corresponding time-periodic Floquet modes in the 
  $(P_1,P_2)$ representation are given by 
%%%%%%%%%%%%%%%%%%%%%%%%%%%%%%%  
\begin{equation}\label{wf2}
\Psi_{\mu,k } (P) = \exp\{- i \sin\left(2 \omega t \right) F(\omega) \} \phi_\mu(P_1) \delta(P_2 - k) \, , 
\end{equation}
the  index $k$ labels the degeneracy of the Landau-Floquet states, and  $\phi_{\mu}\fues{P_1}$ is the harmonic oscillator function  in the 
$P_1$ representation
%%%%%%%%%%%%%%%%%%%%%%%%%%%%%%%
\begin{equation}\label{hofun}
\phi_{\mu}\fues{P_1}=\brak{P_1}{\mu}=
\frac{1}{\sqrt{\pi^{1/2}2^{\mu}\mu !}}
e^{-P_1^2/2}H_{\mu}\fues{P_1} \, , 
\end{equation}
 $H_{\mu}\fues{P_1}$  is the Hermite polynomial and the function $F(\omega)$ is given as 
%%%%%%%%%%%%%%%%%%%%%%%%%%%%%%%
\begin{equation}
\hskip-2.5cm F(\omega) = \frac{\omega_c}{\omega}\left(\frac{e E_\omega l_B }{\omega^2 - \omega_c^2 } \right)^2 \left[  \omega^2 - \omega_c^2  + 2 \omega^2 \epsilon_x^2
- 2 \omega_c^2  \epsilon_y ^2 + \frac{Re(\epsilon_x^* \epsilon_y) }{\omega \omega_c}  \left(2 \omega^4 - \omega^2  \omega_c^2 
+ \omega_c^4 \right)  \right]. 
\end{equation}
 %%%%%%%%%%%%%%%%%%%%%%%%%%

Let us now consider the complete Hamiltonian including the contribution from the periodic potential. 
 When  the transformation induced by  $W(t)$ is applied,   the   Schr\"odinger equation in (\ref{ecs1})  becomes
 %%%%%%%%%%%%%%%%%%%%%%%%%%
\begin{equation}\label{ecs2}
P_0 \Psi^{(W)}  =   H_0 \Psi^{(W)} +   V_W (t) \Psi^{(W)} \, , 
\end{equation}
where $\Psi^{(W)}= W(t)  \Psi$ and  $V_W  (t)  = W(t)  V (\mathbf{r}) W^{-1}(t)$.      Notice that the periodic  potential acquires a time dependence  brought by   the  $W(t)$ transformation. The problem is now solved in the interaction representation  using first order time dependent perturbation theory. 
In the interaction representation $\Psi^{(W)}_I = \exp\{i H_0 t\} \Psi^{(W)}$, and the  Schr\"odinger equation becomes 
 %%%%%%%%%%%%%%%%%%%%%%%%%%
\begin{equation}\label{ecs3}
i \partial_t  \Psi^{(W)}_I  =  \left \{ V_W (t) \right \}_I  \Psi^{(W)}_I  \, . 
\end{equation}
The equation has the solution  $\Psi^{(W)}_I  (t) = U(t - t_0) \Psi^{(W)}_I  (t_0)$, where $U(t)$  is the evolution operator. To first order in perturbation theory it  is given by the expression 
 %%%%%%%%%%%%%%%%%%%%%%%%%%
\begin{equation}\label{opu}
U(t)  = 1 - i \int_{-\infty}^{t}  dt^\prime \left[ W^{\dagger}(t^\prime) V (\mathbf{r}) W(t^\prime)   \right]_I  \, .
\end{equation}
 The interaction is adiabatically  turned off as $t_0 \to  - \infty$,
in which case the asymptotic  state is selected as  one of the Landau-Floquet  eigenvalues of $H_0$, $i.e.$ 
$\vert  \Psi^{(W)}_I (t_0) \rangle \to  \vert  \mu,k \rangle$.
Utilizing the explicit expression for the $W$ transformation in (\ref{opw}) and after a lengthly calculation the matrix elements of 
the evolution operator can be worked out as   
  \begin{equation}\label{meU}
\hskip-2.5cm  {\brakete{ \mu,k}{U(t)  }{ \nu,k^\prime}} = \delta_{\mu\nu} \delta_{kk^\prime} - \sum_l \sum_{mn} 
   \delta(k - k^\prime + l_B q_n^{(y)}
  \frac{e^{il_B q_m^{(x)} (k+l_Bq_n^{(y)}/2)} e^{i \left( {\cal E} _{\mu\nu} + \omega l \right)t}}{{\cal E} _{\mu\nu} + \omega l   + \omega_E  } \, C^{(l)}_{\mu\nu,mn},
\end{equation}
  where $\omega_E = e l_B^2 (q^{(y)}_n E^c_x - q^{(x)}_m E^c_y)$, and the discreet pseudomomentum are given as $ q^{(x)}_m = 2 \pi m /a$ and  $ q^{(y)}_n= 2 \pi n /b$. The explicit expression for $ C^{(l)}_{\mu\nu,mn}$ is  given by  
 %%%%%%%%%%%%%%%% %%%%%%%%%%%%%%%%%%%%%%%%%%
\begin{equation} \label{defc}
 C_{\mu,\nu,mn}^{(l)}   =  \, \frac{ 1}{l_B^2}   \,    V_{m \, n} \,   D_{\mu\nu}(\tilde{q}_{mn} ) \,  \left(\frac{\Delta_{m \,n} }{i \vert  \Delta_{m \, n}  \vert} \right)^l \, 
 J_l  \left( \vert  \Delta_{m \, n}  \vert \right) \, , 
\end{equation}
 where $\tilde{q}_{mn} = i l_B (q^{(x)}_m - i q^{(y)}_n )/ \sqrt{2}$,    $J_l $  denote  the   Legendre polynomials  and   $D_{\mu \nu} $ is given in terms of the  generalized  Laguerre polynomials according to 
\begin{equation}\label{laguerres}
\hskip-1.0cm D^{\nu \mu}\fues{ \tilde{q}}=\braket{\nu}{ D\fues{ \tilde{q}}}{\mu}
=e^{-\frac{1}{2}\abs{  \tilde{q}}^2}
\llal{\begin{array}{ll}\fues{- \tilde{q}^{*}}^{\mu-\nu}
\sqrt{\frac{\nu!}{\mu!}}L^{\mu-\nu}_{\nu}
\fues{\abs{ \tilde{q}}^2}, \,\,\,\,& \mu >\nu, \\
 \tilde{q}^{\nu-\mu}\sqrt{\frac{\mu!}{\nu!}}
L^{\nu-\mu}_{\mu}\fues{\abs{ \tilde{q}}^2},
& \mu <\nu,\\
\end{array}}
\end{equation}
and 
\begin{equation} \label{defdelta2}
\Delta_{m \, n}  =  \frac{\omega_c l_B^2 e E_\omega }{\omega \left( \omega^2 - \omega_c^2 + i \omega \Gamma_{rad}  \right)}
\left[ \omega \left(q^{(x)}_m e_x + q^{(y)}_n e_y  \right)  + i \omega_c \left(q^{(x)}_m e_y  -  q^{(y)}_n e_x  \right) \right] \, . 
\end{equation}
%%%%^^^^^^^^***********%%%%%%%%%%%%%%##############@@@@@@@
Summarizing,  the solution to the original \Sh equation in \Eq{ecs1} has been achieved 
by means os three successive  transformations:
 %%%%%%%%%%%%%%%%%%%%%%%%%%
\begin{equation}\label{3trans}
\vert  \Psi_{\mu,k} (t) \rangle  = W^\dag  \,  \exp\{-i H_0 t\} \,  U(t - t_0) \,  \vert  \mu ,k \rangle ,
\end{equation}
 %%%%%%%%%%%%%%%%%%%%%%%%%%
 the  explicitly  expressions for $H_0$, $W$, and $U$ are given in Eqs. (\ref{tran1}), (\ref{opw}), and (\ref{meU}) respectively. 

 \section{Kubo formula for Floquet states.}\label{kubo}

The usual Kubo formula for the conductivity must be modified in order to include the Floquet dynamics.
In the presence of an additional $DC$ electric field  the complete Hamiltonian is 
 $H_T = H + V_{ext},$
where $H$ is the Hamiltonian in \Eq{ecs1} and $V_{ext} = \frac{1}{m} {\mathbf{\Pi} } \cdot  \mathbf{A}_{ext}$, with  $\mathbf{A}_{ext}= \frac{ \mathbf{E}_0}{\omega} \, sin \left( \Omega t\right) \, exp \left( - \eta \vert t \vert  \right).$ 
The static limit is obtained with $\Omega \to 0$, and $\eta $ represents the rate at which the perturbation is turned on and off. In order to calculate the expectation value of the current density, we need the density matrix  $\rho(t)$  which obeys  the von Neumann equation 
\begin{equation}\label{vn1}
i \hbar \frac{\partial \rho  }{\partial t}  =  \yav{H_T , \rho} =  \yav{H + V_{ext}  , \rho}.
 \end{equation}
We write to first order   $\rho = \rho_0 + \Delta \rho$, where the leading term 
 satisfies the equation  
 \begin{equation}\label{vn0}
  i \hbar \frac{\partial \rho_0  }{\partial t}  =  \yav{H , \rho_0} \, . 
  \end{equation}
In agreement  with  \Eq{3trans},      $\Delta \rho$ is transformed to 
\begin{equation} \label{dmtran}
\hskip-1.5cm  \tilde {\Delta \rho}  (t)   = U^\dag_I(t-t_0)   \exp\{i H_0 t\}  W(t)  \, {\Delta \rho(t)} \,  W^\dag (t)  \exp\{-i H_0 t\}   U_I(t-t_0) \, . 
\end{equation}
In terms of the transformed  density matrix $  \tilde {\Delta \rho}  (t) $,  \Eq{vn1} becomes
%%%%%%%%%%%%%%%%%%%%%%%
 \begin{equation}\label{vn2}
  i \hbar \frac{\partial   \tilde {\Delta \rho}  }{\partial t}  =  \yav{\tilde{V}_{ext} , \tilde{ \rho}_0 },  
   \end{equation}
where $\tilde{V}_{ext}$ and $ \tilde{ \rho}_0$ are the    external potential and quasi-equilibrium density matrix 
transformed  in the same manner as  $\tilde {\Delta \rho}$ in  \Eq{dmtran}. The transformed  
quasi-equilibrium density matrix is assumed to have the form  
$ \tilde{ \rho}_0 = \sum_\mu \vert \mu \rangle f({\cal E}_\mu) \langle \mu \vert ,$ where $f({\cal E}_\mu)$  is the usual Fermi function and ${\cal E}_\mu$ the Landau-Floquet levels. The argument  behind this selection  is an adiabatic assumption 
that the original   Hamiltonian $H$ produces a quasi equilibrium state 
characterized by the  Landau-Floquet eigenvalues \cite{tor1}.
It is straightforward to verify that  this selection guarantees that the 
quasi-equilibrium condition in  (\ref{vn0}) is verified.  Using the results in Eqs. (\ref{3trans})  and   (\ref{dmtran}), 
the  expectation value  of the density matrix   can now  be  easily obtained from the integration of \Eq{vn2} 
  with the initial  condition $ {\Delta \rho}  (t)  \to 0$ as $t \to -\infty$ giving for $t < 0$ 
 %%%%%%%%%%%%%%%%%%%%%%%%%%
\begin{eqnarray}\label{emdm}
\hskip-2.2cm  {\brakete{\Psi_{\mu,k}}{ {  {\Delta \rho}  (t) } }{\Psi_{\nu,k^\prime}}}     &=&
  {\brakete{\mu,k}{ { \tilde {\Delta \rho}  (t) } }{\nu,k^\prime}}        \nonumber \\
&=&  \frac{e \mathbf{E}_0}{2} \cdot  \int_{-\infty}^t \left[    
  \frac{e^{i(\Omega - i \eta) t^\prime}}{\Omega}  f_{\mu \nu}  {\brakete{\Psi_{\mu,k}}{  \mathbf{\Pi} (t^\prime)}{\Psi_{\nu,k^\prime}}}  + \left(\Omega \to - \Omega \right) \right],
\end{eqnarray}
where  the definition  $f_{\mu \nu} =  f({\cal E}_\mu) - f({\cal E}_\nu)$ was used.
The  expectation value for the momentum operator  is explicitly computed with the help  of Eqs   (\ref{opw}),  (\ref{meU}), and   (\ref{3trans}), after a lengthly calculations it yields 
 %%%%%%%%%%%%%%%%%%%%%%%%%%
  \begin{equation}\label{emopm}  
\hskip-2.5cm   {\brakete{\Psi_{\mu k}}{ \mathbf{\Pi}_i}{\Psi_{\nu k^\prime}}} =    \, \sqrt{e B} \,  \sum_l \sum_{mn} \, 
   \delta(k - k^\prime + l_B q_n^{(y)}) e^{il_B q_m^{(x)} (k+l_Bq_n^{(y)}/2)} \,  e^{i ( {\cal E} _{\mu \nu} + \omega l - i \eta  t)} \, \, 
  \Delta ^{(l)}_{\mu \nu,mn} (j) .
 \end{equation} 
 %%%%%%%%%%%%%%%%%%%%%%%%%%
Here the following definitions were introduced:  $ {\cal E} _{\mu \nu} =  {\cal E} _\mu - {\cal E} _\nu$,  $a_j = b_j= 1 $ if $j=x$ and $a_j = -b_j= -i $  if $j=y$,  and   $ \Delta ^{(l)}_{\mu \nu,mn} (j)$ is given by 
 %%%%%%%%%%%%%%%%%%%%%%%%%%
\begin{equation} \label{defdelta}
  \Delta ^{(l)}_{\mu \nu,mn} (j) =
  -  \frac{1}{\sqrt{2}} \left[  \frac{  a_j \tilde{q}_{mn}^* C_{\mu\nu,mn}^{(l)}}{{\cal E} _{\mu\nu} - \omega_c + \omega l - i \eta }
 +   \frac{ b_j \tilde{q}_{mn} C_{\mu\nu,mn}^{(l)}}{{\cal E} _{\mu\nu} + \omega_c + \omega l - i \eta }\right],
\end{equation}
the expression for $C^{(l)}_{\mu\nu,mn}$ are given in (\ref{defc}).
Utilizing these results the time  integral in \Eq{emdm} is readily carried out. 
 The current density to first order in the external electric field  can now be calculated from 
  $\langle  \mathbf{J} (t,  \mathbf{r}) \rangle  = Tr \left[   \tilde {\Delta \rho}  (t)   \tilde \mathbf{J} (t)  \right] $, the resulting expression represents the local      density current. Here we are concerned with the macroscopic  conductivity tensor that relates the spatially and time averaged  current density  
  $  \mathbf{j} = \left( \tau_\omega {\cal A}  \right)^{-1} \int_0^{\tau_\omega} dt \int d^2x \langle  \mathbf{J} (t,  \mathbf{r}  ) \rangle$ to the averaged electric field; here $\tau_\omega = 2 \pi / \omega$ and ${\cal A}$ is the area of the system (it is understood that ${\cal A} \to \infty)$.
  Assuming that the external electric field points along the $x$-axis the macroscopic conductivity  can be worked out,  results for  the dark and microwave induced  conductivities  are quoted:
   %%%%%%%%%%%%%%%%%%%%%%%%%%  
\begin{equation}\label{condd}
\hskip-1.5cm \mathbf{\sigma}^D_{xi}  =  i \frac{e^2 \omega_c^2}{4 \hbar l_B^2} \sum_{\mu \nu} \left\{    \frac{ f_{\mu\nu}}{\Omega}
  \left[ \frac {a_i \mu \delta_{\mu , \nu +1} } {{\cal E}_{\mu\nu} +\Omega - i \eta } 
  + \frac {b_i \nu \delta_{\mu , \nu -1} } {{\cal E}_{\mu\nu} +\Omega - i \eta }  \right]        + \left(\Omega \to - \Omega \right)   \right\},
 \end{equation}
 %%%%%%%%%%%%%%%%%%%%%%%%%%
\begin{equation}\label{condw}
\hskip-1.5cm \mathbf{\sigma}^\omega_{xi}    =   \frac{e^2 \omega_c^2}{4 \hbar} \sum_{\mu \nu} \left\{    \frac{ f_{\mu\nu}}{\Omega}
 \sum_{mn}  \sum_l \frac{\Delta_{\mu\nu,mn}^{(l)} (i) \Delta_{\nu\mu,mn}^{(-l) }(x) }{ {\cal E}_{\mu\nu} + \omega l + \Omega - i \eta}
   + \left(\Omega \to - \Omega \right) \right\}.
 \end{equation}
  In these expressions   the external electric field points along the $x$-axis.
 Hence, setting $i =x$ or $i=y$  the 
 longitudinal and Hall conductivities  can be selected. The denominators on the R.H.S. of the previous equations 
 can be related to the advanced and  retarded  Green's functions   $G^{\pm}_\mu ({\cal E} ) = 1/\left({\cal E}  -{\cal E} _\mu  \pm i \eta  \right) $. To make further progress the real and absorptive parts of the 
 Green's functions are separated  taking the limit  $\eta \to 0$  and using 
 $lim_{\eta \to 0} \, 1/({\cal E}  -  i \eta) = P 1/{\cal E}   + i \pi \delta({\cal E} )$, where $P$ indicates the principal-value integral. As
 usual  the real and imaginary  parts contribute to the Hall and  longitudinal conductivities respectively.
  However, the  previous expression  would present a singular  behavior that is  an artifact of the $\eta \to 0$ limit. This problem is solved by including the  disorder broadening effects.
    A formal procedure requires to  calculate   the broadening produced by the disorder potential, this calculation has been   carried out  by Ando \cite{ando1} and 
    Gerhardts \cite{gerh} within the Born Approximation;  the density of states for the $\mu$-Landau level   can be represented by  a Gaussian-type form \cite{ando2}
 \begin{equation} \label{denst}
\hskip-1.0cm  Im \, G_\mu({\cal E} ) = \sqrt{\frac{\pi}{2 \Gamma_\mu^2}}  \exp{\left[ - ({\cal E}  - {\cal E} _\mu)^2 /(2 \Gamma_\mu^2) \right] } , 
 \hskip1.5cm    \Gamma^2_\mu = \frac{ 2\beta_\mu  \hbar^2 \omega_c  }{ (\pi \tau_{tr})} .
 \end{equation}   
The parameter  $\beta_\mu $ in  the level width  takes into account the difference  of the transport scattering time $\tau_{tr}$  determining the mobility $\mu$,  from the single-particle  lifetime $\tau_s$. In the case of short-range scatterers  $\tau_{tr} = \tau_s$ and  $\beta_\mu =1 $. An expression for $\beta_\mu$,  suitable for numerical evaluation,   that applies for a long-range screened potential  is given in
reference \cite{tor1};  $\beta_\mu$  decreases for higher Landau levels; $e.g.$ $\beta_0 \approx 50$, $\beta_{50}  \approx 10.5$.

The static  limit with respect to the external field is obtained taking 
   $\Omega \to 0$   in \Eqs{condd} and (\ref{condw}).  In what follows  results are presented  for the   longitudinal microwave induced conductivity, the corresponding dark conductivity expressions as well as the Hall microwave induced conductance  are quoted  in the appendix. The final result 
 for the  microwave induced longitudinal conductance is worked out as  
  %%%%%%%%%%%%%%%%%%%%%%%%%%
 \begin{equation}\label{condLw}
\hskip-2.5cm  \mathbf{\sigma}_{xx} ^\omega   =   \frac{ e^2 }{\pi \hbar l_B^2}  
\int d {\cal E}  \sum_{\mu \nu}  \sum_l  \,   \sum_{m \, n}  \, Im G_\mu \left({\cal E}   \right)B^{(l)}   \left({\cal E}  ,{\cal E} _\nu \right)       \,
  \bigg{ \vert} q_n^{(y)} \,  J_l\left( \vert \Delta_{m \, n}  \vert \right)  V_{m \, n}   D_{\mu\nu} (\tilde{q}_{mn}) \bigg{\vert}^2  ,
 \end{equation}
 where the following  functions have been defined   
 \begin{equation} \label{derspec} 
\hskip-2.0cm  B^{(l)} ( {\cal E} ,  {\cal E} _\nu) =  -  \frac{d}{d{\cal E} _0}  \bigg{ \{ } \left[ f( {\cal E} + l \omega + \omega_E + {\cal E} _0)-  f ( {\cal E} )\right]  
 Im \, G_\nu ( {\cal E}  + l \omega  + \omega_E  + {\cal E} _0) \bigg{ \}}
  \bigg{\vert}_{{\cal E} _0= 0}. 
 \end{equation}

%%%%%%%%%%%%%%%%%%%%%%%%%%%%%%%%%%%%%%%%%%%%%%%%%%%
 \section{Results.}\label{results}
 The expression in \Eq{condLw} can be numerically evaluated after the  Fourier components $ V_{m \, n}$ of the periodic  potential    are  specified.   We consider two examples, a square lattice potential of the form 
   \begin{equation}  \label{pote1} 
  V(\mathbf{r}) = V_0 \left[ \cos\left( \frac{2 \pi x }{a} \right)  +  \cos\left( \frac{2 \pi y }{a} \right) \right]  \, , 
 \end{equation} 
 and a hexagonal potential given by \cite{clarow}

 \begin{equation}  \label{pote2} 
\hskip-2.0cm  V( \mathbf{r}) = V_0 \left[  \cos\left\{  \frac{2 \pi  }{a} \left( \frac{x}{\sqrt{3}} + y \right) \right \}   + 
   \cos\left\{  \frac{2 \pi }{a} \left( \frac{x}{\sqrt{3}} - y \right) \right \}  + \cos \left \{ \frac{4 \pi x }{\sqrt{3} a }  \right \} \right] \, . 
 \end{equation}

In our calculations it is assumed  that  a  lateral  superlattice  is cleaved   in ultraclean  $GaAs/Al_xGa_{1-x} As$ sample 
with high  electron  mobility,  $ \mu \sim 0.5-2.5 \times 10^7 cm^2/V  s$; the periodic  potential  has the form given in Eqs. (\ref{pote1},\ref{pote2}) with parameters in the   range   $a \sim 50 -200 \, nm$  and  $V_0=1 \, meV $. The other parameters  of the sample 
 are estimated as  effective  electron mass $m^* = 0.067 \, m_e$,       fermi energy $\epsilon_F = 10 \, meV$, electron density $n = 3 \times 10^{11} cm^{-2}$,  and temperature $T =1 \,  K$.  For  the applied external field we consider   magnetic fields in the range $0.05 - 0.4 \, \, Tesla$,  microwave frequencies $f  \sim 10-200 \, Ghz$,  
and ac-electric field intensity   $ \vert \vec E_\omega \vert  \sim 10-100 \, V/cm$ that corresponds to a microwave power  characterized by the dimensionless quantity  $\alpha \sim \frac{c \epsilon_0 \vert  E_\omega\vert ^2}{m^* \omega^3}$ that varies in the range $\alpha \sim 0.2-1000$.  The relaxation time $\tau_{tr}$  in \Eq{denst}  is related with  the zero field  electron mobility  through 
   $\mu = e \tau_{tr} /m^*$,  and $\beta_\mu  \approx 10.5$, a value that is justified for large filling factors   $\mu \approx 50$ \cite{tor1}.  A detailed account of the electron dynamics requires to distinguish  between  various 
   time life's; following  reference \cite{mikha},   $\Gamma_{rad}$ in \Eq{soleqm}  is related to 
   the   radiative decay  width   that is interpreted as coherent dipole re-radiation of  electromagnetic waves by the oscillating 2D
   electrons excited by microwaves. Hence, it is given by 
    $\Gamma_{rad} = 2 \pi^2 \hbar n e^2 /\left( 3 \epsilon_0 c \, m^* \right)$,  using the values of $n$ and $m^*$ given above it yields  $ \Gamma_{rad} \sim 2.2  \, meV $.  In all the  examples, except in    \Fig{figure6}, we consider the linear regime, 
     the dc-electric field will be included only through the Kubo formula, hence  
    $ \omega_E = 0$ in Eqs. (\ref{condLw}), (\ref{derspec}). In the case of  \Fig{figure6} the non-linear dc-electric field  effects are  included 
  using the solution to the classical equations of motion  with both ac- and dc-electric fields  (\ref{soleqm}).

   Adding the dark and microwave  induced  conductivities, the total longitudinal $\sigma_{xx} = \sigma_{xx}^D +
    \sigma_{xx}^w  $, and Hall $\sigma_{xy} = \sigma_{xy}^D +
    \sigma_{xy}^w  $ conductivities are obtained. It should be pointed out that the interference between the dark and microwave contributions exactly cancels. The corresponding resistivities are obtained from the expression 
   $ \rho_{xx} =\sigma_{xx}/\left(  \sigma_{xx}^2 + \sigma_{xy}^2 \right)$ and  $ \rho_{xy} =\sigma_{xy}/ \left( \sigma_{xx}^2 + \sigma_{xy}^2 \right)$. The relation $\sigma_{xy} \gg \sigma_{xx}$ holds in general,  hence it follows that $\rho_{xx} \propto \sigma_{xx}$,  and the longitudinal resistivity  follows the same oscillation  pattern as  that of  $\sigma_{xx}$.

\begin{figure} [hbt]
\begin{center}
\includegraphics[width=4.5in]{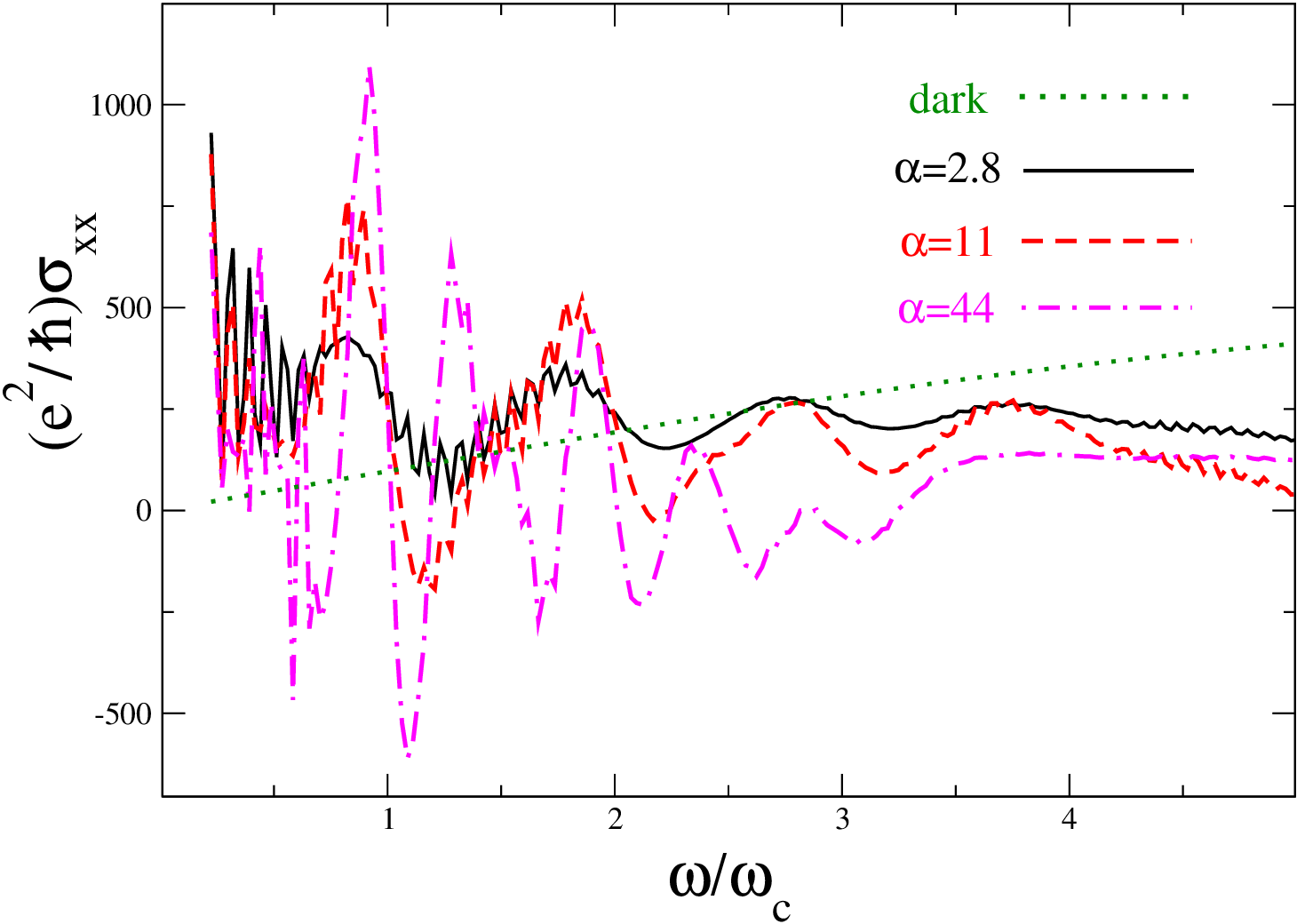}
\end{center}
%\vskip0.5cm
\caption{ Longitudinal conductivity   as a function $\epsilon =\omega/\omega_c$ for four  values of the microwave power intensity. The dotted line
corresponds to the dark case (without microwave radiation) the other plots are for: 
$\alpha = 2.8$  continuos line; $\alpha = 11$  dashed line;  and $\alpha = 44$  dashed-dotted line. The microwave  polarization  is linear transverse (with respect to the current),  the  frequency $f  = 25  \, Ghz$   and the   power is characterized by the dimensionless parameter 
$\alpha = c \epsilon_0 \vert E_\omega \vert^2/(m^* \, \omega^3)$. The potential corresponds to a square 
lattice, \Eq{pote1}, with $a = 100 \, nm$, and $V_0=1 \, meV $.  The other parameters are selected as follows:
$m^* = 0.067 \, m_e$,  $ \mu \approx 2.5 \times 10^7 cm^2/V  s$,   
$\epsilon_F = 10 \, meV$, $T =1  \,  K$.  
}
\label{figure1}
\end{figure}

   \Fig{figure1} displays plots of the total longitudinal conductivity  as a function of   $\epsilon = \omega/ \omega_c  $. 
   Whereas the dark conductivity presents the expected linear behavior, the total longitudinal conductance  shows a strong oscillatory behavior,    
   with distinctive negative conductance  states. The periodicity as well as the  number of negative conductance regions depends on the 
intensity of the   microwave radiation. In the cases $\alpha =2.8 $ and $\alpha = 11$ we observe $\epsilon$-periodic oscillations with minima centered
at    $\epsilon =1.1$ , $\epsilon = 2.1$, and  $\epsilon =3.1$. Notice that  only for the $\alpha = 11$ case the minima correspond to NCS. 
A further increase in the microwave intensity ($\alpha = 44$) yields  several negative conductance states.  The oscillation period is now reduced to 
$\frac{1}{2}  $.
In general  it is observed  that $\sigma_{xx}$ vanishes at $\omega /\omega_c =j$ for $j$ integer.   The oscillations follow a pattern with minima
centered  at $\omega/\omega_c =j + \frac{1}{2} (l-1) +  \delta $,  and maxima centered at  $\omega/\omega_c =j + \frac{1}{2} (l-1) -  \delta $, where $j=1,2,3.......$,   $\delta \approx 1/5 $ is a constant phase shift and $l$ is the dominant multipole that contributes to the conductivity in \Eq{condLw} . For moderate microwave  power the  $l=1$ ``one photon'' stimulated  processes dominate, corresponding to what is observed for  $\alpha =2.8 $ and $\alpha = 11$ . In the last  example  ($\alpha = 44$), the results can be interpreted as the results of  ``two  photon''  processes 
($l =2$).
For small microwave power, $\sigma_{xx}^\omega$ is dominated by  $l=0$ Bessel term in \Eq{condLw}, that
  is always positive. Negative conductance states arise when the    $l=1$ and   $l=0$ terms  become 
comparable:  $  J_0\left( \vert \Delta  \vert \right)  B^{(0)}  \sim  J_1\left( \vert \Delta  \vert \right)  B^{(1)}$. A simple analysis 
show that this condition is fulfilled for $ \vert \Delta  \vert \sim  0.1$. Using the result in \Eq{defdelta2},  the condition to produce 
NCS can be estimated  as $ \vert \mathbf{E}_\omega \vert  > E_{th}$ where $ E_{th} \approx 0.1\,  a \,  \Gamma_{rad} / \sqrt{8} e l_B $.  
For the parameter used in   \Fig{figure1},    $ E_{th} \approx 10 \, V /cm $ or $\alpha_{th}  \approx 10$,  in good agreement with the results 
displayed  by the plots.  

\begin{figure} [hbt]
\begin{center}
\includegraphics[width=4.5in]{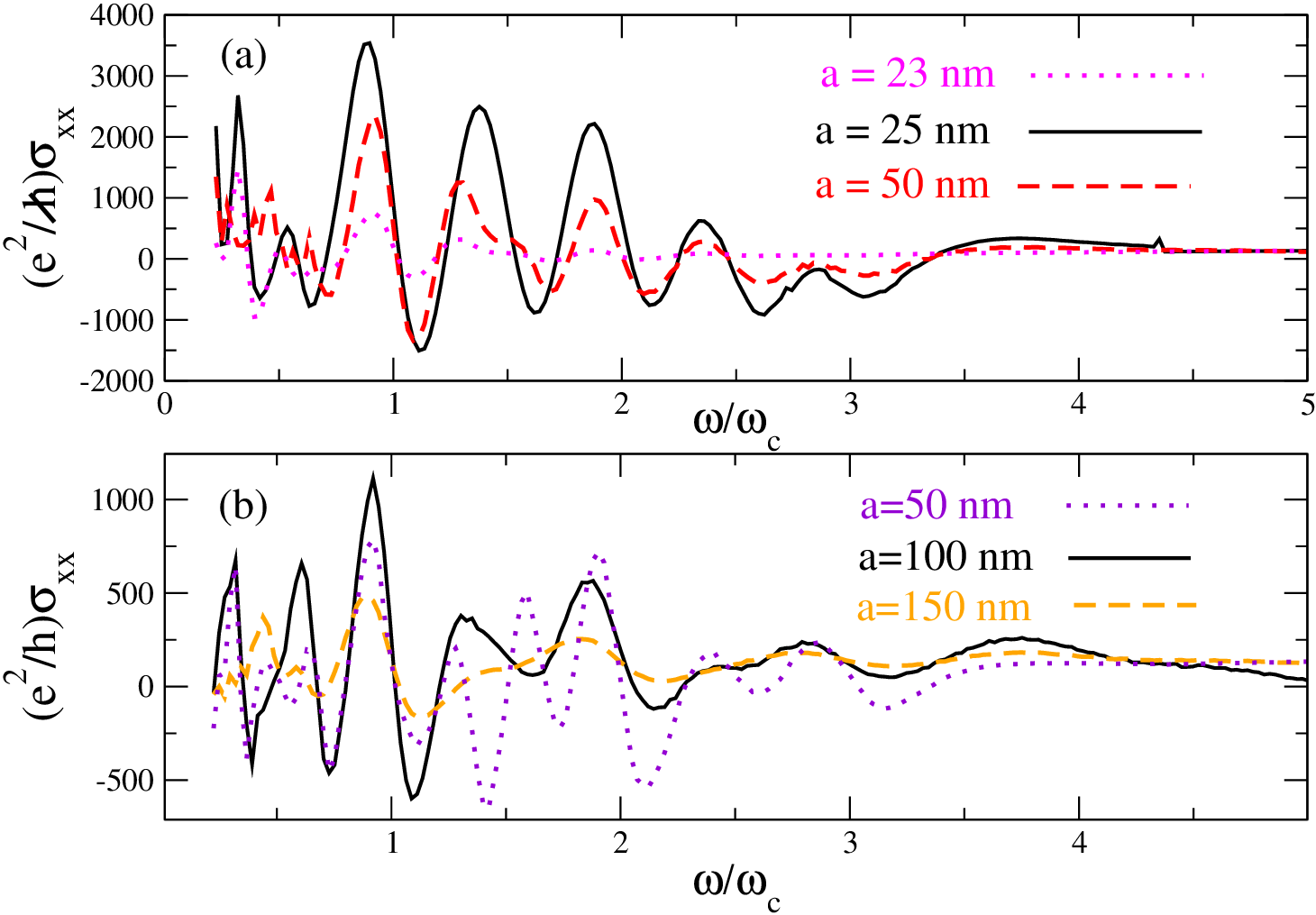}
\end{center}
%\vskip0.5cm
\caption{  Longitudinal conductivity   as a function  of $\epsilon =\omega/\omega_c$ for:  (a) Square lattice, \Eq{pote1}: $a=23 \, nm$ dotted line,  $a=25 \, nm$  continuos line,  and  $a=50 \, nm$    dashed
  line. (b) Hexagonal  lattice, \Eq{pote2}: $ a=50 \, nm$ dotted line,  $a=100 \, nm$    continuos line,  and 
   $a=150 \, nm$ dashed-dotted line. The microwave power is given by  $\alpha =  11$,
  corresponding to an ac-electric field intensity of $  E_\omega =10 \, V/cm$. The other parameters 
  have the same values as in   \Fig{figure1}.  
  }
\label{figure2}
\end{figure}

Next we consider the dependence of $\sigma_{xx}$ on the lattice parameter $a$.   Plots of  $\sigma_{xx}$
versus  $\epsilon = \omega/ \omega_c  $ for various selections  of $a$  are presented  in the case of a rectangular   (\Fig{figure2}a)
and hexagonal lattice    (\Fig{figure2}b).  A value of $a_{max}$ for which  the oscillation amplitude of $\sigma_{xx}$  attains  its maximum value
can be identified in both cases. The precise value depends on the selection of the other parameters. In the present case: 
$a_{max}  \approx 25 \, nm$  for the  rectangular lattice  (\Fig{figure2}a), whereas $a_{max} \approx 100 \, nm$ for the  hexagonal  lattice  (\Fig{figure2}b).  These plots clearly show that the oscillatory behavior of $\sigma_{xx}$ and in particular the regions of negative 
conductivity are   governed by the ratio $\epsilon = \omega /\omega_c$.

Negative conductivity requires ultra-clean  samples,  the phenomena appears when the 
     electron mobility exceeds  a threshold  $ \mu_{th}$.  \Fig{figure3}  displays $\sigma_{xx} \, \, v.s. \,\, \omega / \omega_c $ plots for three  selected values of $\mu$. For $\mu \approx 0.5  \times 10^7 cm^2/V  s$  an   almost linear behavior $\sigma_{xx} \propto 1/B$ is clearly depicted.  As the electron mobility  increases to $\mu \approx 1.5  \times 10^7 cm^2/V  s$, the conductivity  oscillations are clearly observed; however,  negative conductance  states only  appear when   the mobility is increased to $\mu \approx 2.5  \times 10^7 cm^2/V  s$. 
         Eqs. (\ref{condLw},\ref{derspec}) contain the main ingredients that explain the huge increase observed in  the longitudinal  conductance (and resistance),   when the material is irradiated by microwaves and its critical dependence on the electron mobility.  In the standard expression for the Kubo formula there are no Floquet  replica contribution, hence $\omega$   can be set to zero in (\ref{derspec}), if that is   the case $B^{(l)}$ becomes proportional  to the energy derivative  of the Fermi distribution, that  in the $T \to 0$ limit  becomes of the form $\delta ({\cal E}  - {\cal E} _F)$, and the conductivity is positive definite  depending  only on  those states lying  at the Fermi level.  On the other hand,  as a result of the periodic structure induced by the  microwave radiation,  $B^{(l)}$  contains a second  contribution proportional to  the derivative of the density of states:   $   \frac{d}{d{\cal E} }  Im \, G_\nu ( {\cal E}  + l \omega )$.   Due to the oscillatory structure of the density of states, this extra contribution takes both  positive and negative values. According to \Eq{denst} this second term (as compared to the first one) is proportional to  the electron mobility, hence  for sufficiently  high mobility the new contribution dominates leading to  negative conductance states.

   The model can  be used in order to test  chirality effects induced by the magnetic field.    \Fig{figure4}   shows the results  for different 
   $\mathbf{E}_\omega$ field polarization's with respect to the current.   In  (\Fig{figure4}a) it is observed that in the $\epsilon >1$  region,   the  amplitudes of  the $\sigma_{xx}$   oscillation  are  slightly  bigger for transverse  polarization as compared to longitudinal polarization.
   However for  $\epsilon  <1$ a negative conductance region around  $\epsilon = 0.6$ is observed only for longitudinal polarization. 
   A more significant signature is   observable for circular polarization.  Selecting  negative circular polarization (see \Fig{figure4}b), the oscillation amplitudes get the maximum possible value.  Instead,  for positive circular polarization  an important reduction of the oscillatory  of amplitudes  is observed  leading to the  total disappearance  of   the negative conductance states . These results are understood  recalling  that   for negative circular polarization and $ \omega \approx \omega_c$ the  electric field rotates in phase with respect to the  electron cyclotron rotation.

   The present formalism can also be used in order to explore the non-linear regime in which multiple photon exchange  play an essential role. As the microwave radiation  intensity is increased,    higher   multipole  ($l$) terms needs  to be evaluated. In the explored regime   convergent results are obtained  including terms up to the $l=3$ multipole.  In \Fig{figure5}a results are presented   for the longitudinal conductivity as a function of the electric field strength  ($E_\omega$) 
   of the microwave radiation.  Results are displayed for various selections of the magnetic field intensity. 
   In all cases $\sigma_{xx}$ start from a positive value for $E_\omega=0$. For $B=0.067$, the longitudinal conductivity remains positive for all values of the microwave intensity; instead  $B=0.055$ the conductivity remains negative above the threshold  $E_\omega\approx 5  \, V/cm$. 
   On the other hand for the selections $B=0.047$ and  $B=0.084$, $\sigma_{xx}$ displays an oscillatory behavior, with alternating  positive and negative regions.

  The non-linear regime with respect to the dc-external  field can also be explored within the present formalism. The effect is included 
  using the solution to the classical equations of motion  with both ac- and dc-electric fields, \Eq{soleqm}.  
   A possible connection between the observed  $ZRS$ in  $GaAs/Al_xGa_{1-x} As$ heterostructures  \cite{zudov1,mani1,zudov2,mani2} and the  the predicted  NRS\cite{ry1,ry2,durst,andre,shi,lei,vavilov,tor1}   was put forward by Andreev  $etal.$  \cite{andre}, noting that a general analysis of Maxwell equations shows that $NRS$  induces an instability  that  drives  the system  into a  ZRS. This mechanism requires  the longitudinal current $ j _{xx}$  as a function of $E_{dc}$ to have a single minimum, the system  instability will evolve  to the value $E_{dc}$ in which $ j _{xx}$ cancel.  Returning to the 
   irradiated superlattice  case, in \Fig{figure6} 
   it is observed that in general the  $ j _{xx}$ $vs.$  $E_{dc}$ plot has an oscillatory behavior, with more than one minima.
   Hence   the conditions of the Andreeev  mechanism do not apply. Consequently, negative 
   conductance states may be probably observed in 2-dimensional  lateral superlattices, when  exposed to both  magnetic and microwave fields.

 \section{Conclusions.}\label{conclu}

We have considered  a model to describe the conductivity of an electron in a 2-dimensional lateral superlattice subjected to both a magnetic 
field and microwave radiation.  We presented a thorough discussion of the  method  to take into account  the Landau 
and microwave contributions in a non-perturbative exact way, the periodic potential effects are treated perturbatively. The method exploits  the   symmetries  of the problem:  the exact  solution of the  Landau-microwave dynamics (\ref{tran1}) is obtained in terms of the electric-magnetic generators (\ref{trancan})  as well as the solutions  to the  classical  equations of motion (\ref{lagrang}).  The  spectrum  and Floquet modes are explicitly worked out.   In our model,  the Landau-Floquet  states act coherently with respect to the  oscillating field of  the superlattice potential, that   in turn induces  transitions between these levels.  Based on this formalism, a  Kubo-like   formula is provided,  it takes into account the oscillatory Floquet structure of the problem.

\begin{figure} [hbt]
\begin{center}
\includegraphics[width=4.5in]{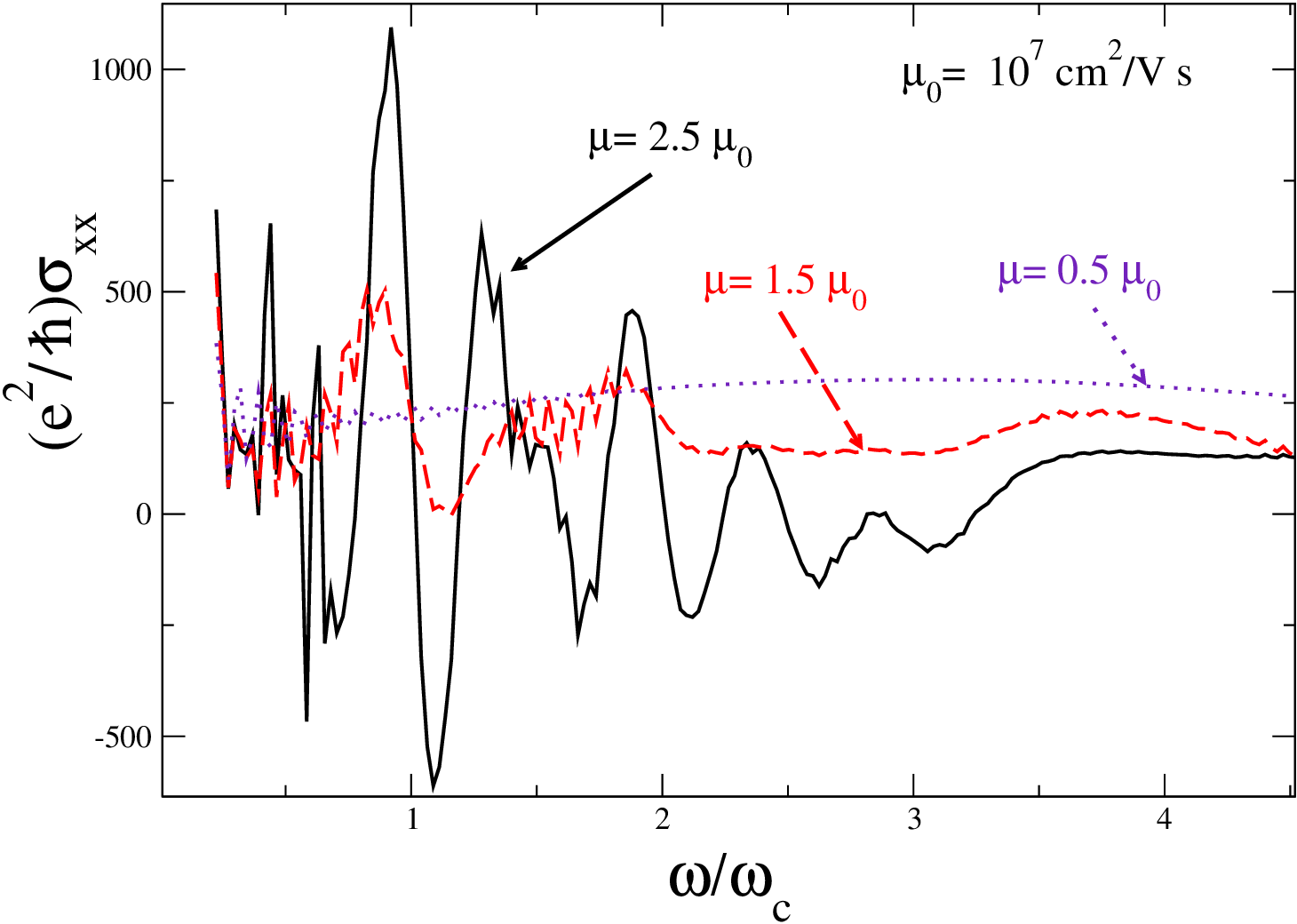}
\end{center}
%\vskip0.5cm
\caption{ Longitudinal conductivity   as a function $\epsilon =\omega/\omega_c$ for three  values
of the electron mobility: $\mu = 0.5 \times 10^7 \, cm^2/Vs  $   dotted line, $\mu = 1.5 \times 10^7 \, cm^2/Vs  $ dashed line, and 
 $\mu = 2.5 \times 10^7 \, cm^2/Vs  $  continuos line. The microwave power is given by  $\alpha =  11$,  the other parameters 
  have the same values as in   \Fig{figure1}. 
 }
\label{figure3}
\end{figure}

\begin{figure} [hbt]
\begin{center}
\includegraphics[width=4.5in]{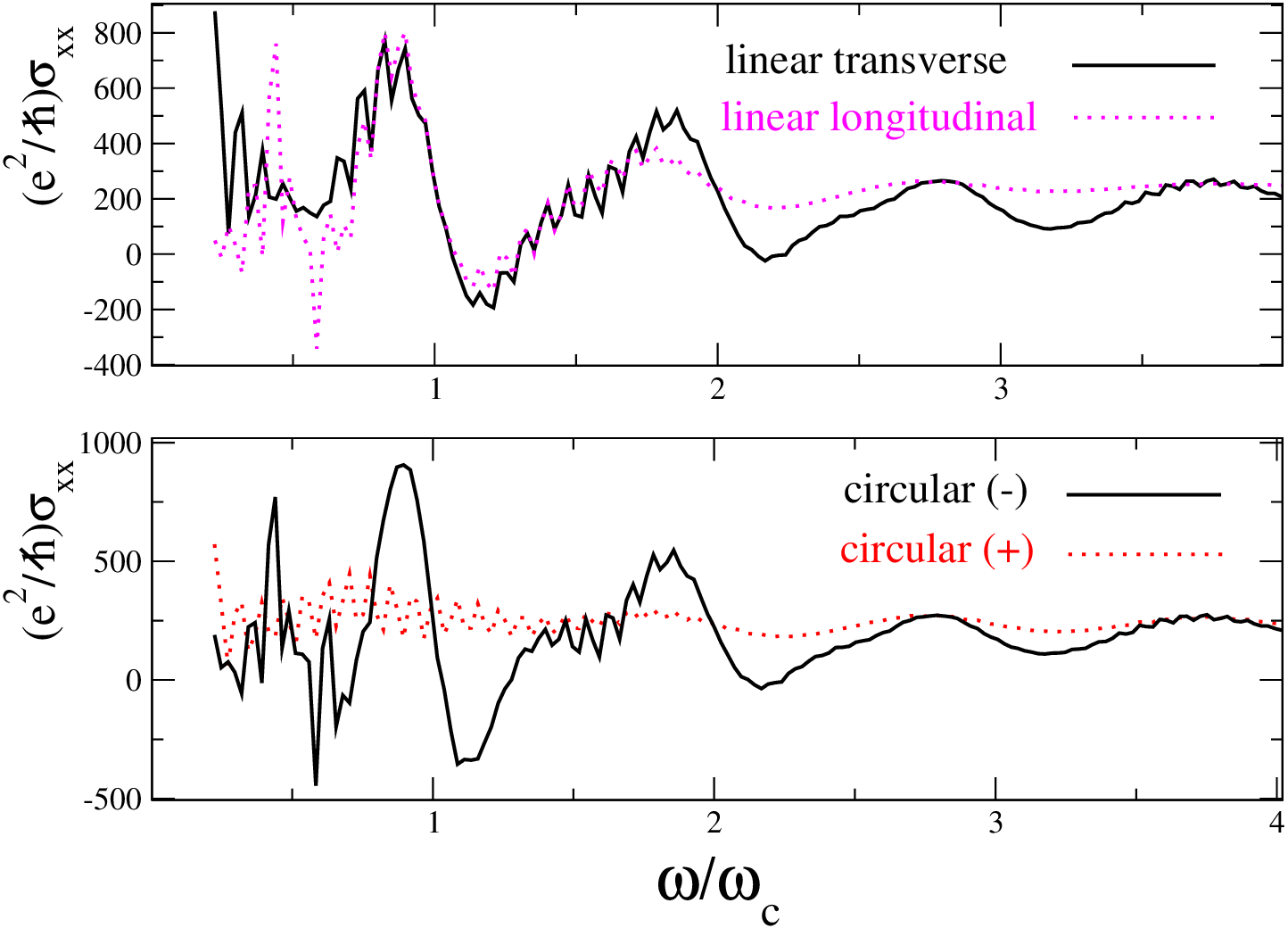}
\end{center}
%\vskip0.5cm
\caption{ Longitudinal conductance $vs.$  $\epsilon =\omega/\omega_c$  for   various   microwave $E_\omega-$field polarization's with respect to the current.  In figure $(a)$ the continuos and dotted  lines correspond to  linear   transverse and longitudinal  polarization's   respectively. 
Figure $(b)$  shows results for circular polarization's:  left-hand (continuos line) and  right-hand  (dotted line). $\alpha =  11$ and the values of the  other  parameters are the same as in figure \Fig{figure1}
 }
\label{figure4}
\end{figure}

\begin{figure} [hbt]
\begin{center}
\includegraphics[width=4.5in]{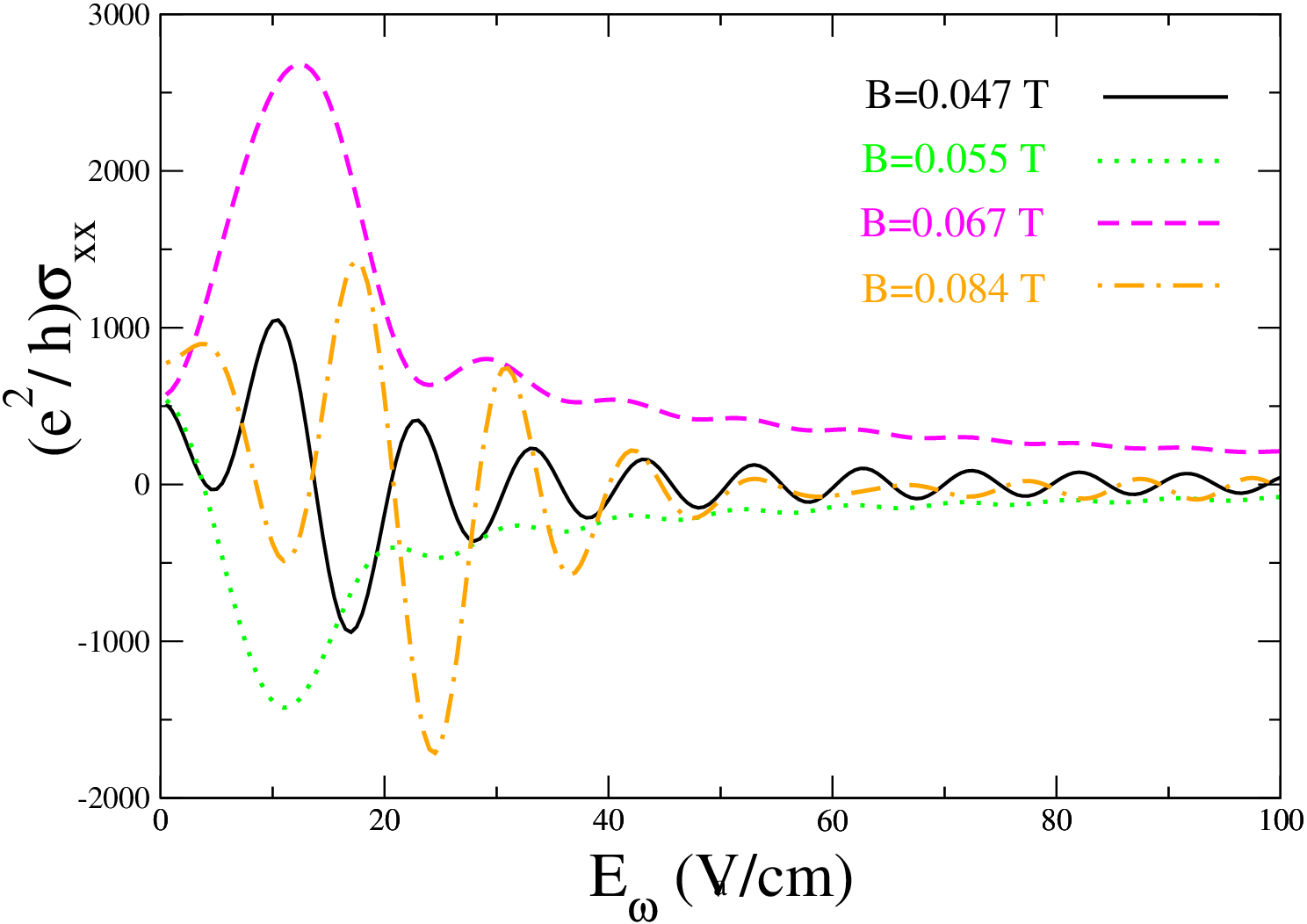}
\end{center}
%\vskip0.5cm
\caption{ Longitudinal conductivity   as a function of the microwave ac-electric field  for four values 
of the magnetic field intensity: 
$B=0.047 \, T $ continuos line, $B=0.055 \, T $ dotted line,   
$B=0.067 \, T $dashed line, and  $B=0.084 \, T $ dashed-dotted  line. 
 The other parameters 
  have the same values as in   \Fig{figure1}
  }
\label{figure5}
\end{figure}

\begin{figure} [hbt]
\begin{center}
\includegraphics[width=4.5in]{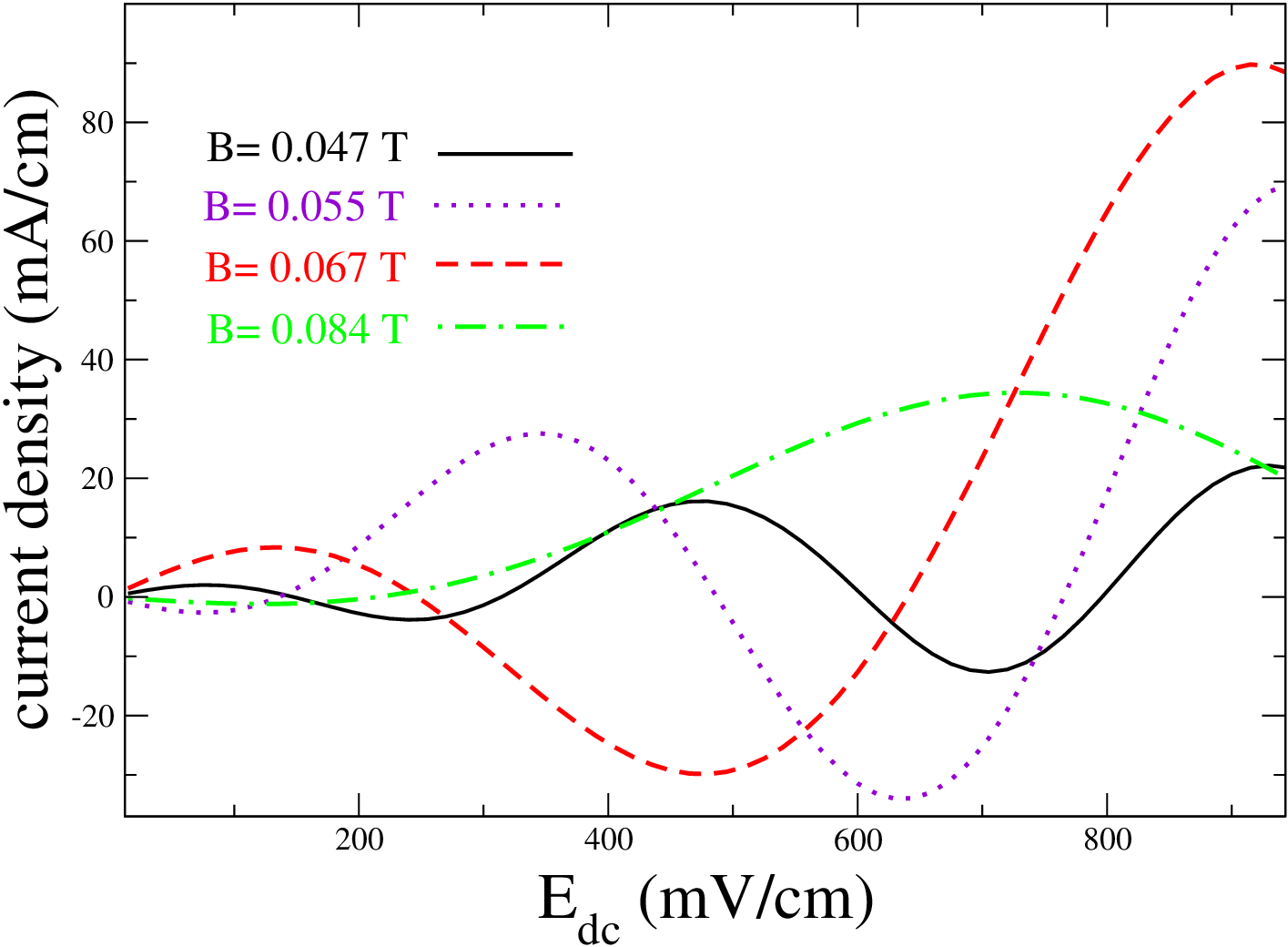}
\end{center}
%\vskip0.5cm
\caption{  Current-voltage characteristics for the irradiated sample 
 for four values 
of the magnetic field intensity: 
$B=0.047 \, T $ continuos line, $B=0.055 \, T $ dotted line,   
$B=0.067 \, T $dashed line, and  $B=0.084 \, T $ dashed-dotted  line. 
 The other parameters 
  have the same values as in   \Fig{figure1}
 }
\label{figure6}
\end{figure}

 It is found  that  $\sigma_{xx}$  exhibits strong oscillations determined by   $\epsilon = \omega /  \omega_c$. 
The oscillations follow a pattern with minima
centered  at $\omega/\omega_c =j + \frac{1}{2} (l-1) +  \delta $,  and maxima centered at  $\omega/\omega_c =j + \frac{1}{2} (l-1) -  \delta $, where $j=1,2,3.......$,   $\delta \approx 1/5$ is a constant phase shift and $l$ is the dominant multipole  contribution. 
 NCS develop for sufficiently  strong  microwave power (\Fig{figure1})  and high  electron mobility 
 (\Fig{figure3}).
According to the Eqs.    (\ref{condLw}) and  (\ref{derspec})  the  longitudinal photoconductivity contains a  new contribution proportional to the derivative of the density of states:
   $   \frac{d}{d{\cal E} }  Im \, G_\nu ( {\cal E}  + l \omega )$.   Due to the oscillatory structure of the density of states this extra contribution takes both  positive and negative values. This  term is proportional to the electron mobility, hence  for sufficiently  high mobility the new contribution dominates leading to  negative conductivity  states.

  An interesting prediction of the present model is related to  polarization effects.  While the results for the cases of linear transverse or longitudinal polarization's show small differences, the selection of circular polarized radiation leads to significant signatures. 
    The maximum possible value  for  the oscillation amplitudes of $\sigma_{xx}$ appears for negative circular polarization.  Instead,   positive circular polarization   yields an important reduction  on  the oscillation   amplitudes   and  the  total disappearance  of   the NCS.  This result can be understood, if one recalls    that for  negative circular polarization and $ \omega \approx \omega_c$ the  electric field rotates in phase with respect to the  electron cyclotron rotation.

In conclusion, it is  proposed that 
 the   combined effect of a perpendicular magnetic field plus the  irradiation of lateral    superlattices can give rise to interesting oscillatory conductance phenomena, with the possible   development of negative conductance states $(NCS)$.  One should stress  that according to our results,  the production of NCS requires ultra-clean samples with electron mobilties of order  $\mu \approx 2.5  \times 10^7 cm^2/V  s$ (see \Fig{figure3}). The electron mobilities 
in the  lateral superlattices fabricated so far  \cite{klitbut} are $\mu \approx  2.5  \times 10^6 cm^2/V  s$,  consequently 
an increase on the electron mobilities  of these kind of experimental setups  by an order of magnitude would be required in order to observe NCS.

%%% Acknowledgments %%%  
\ack 
 We acknowledge the partial financial support endowed by
CONACyT through grants No.   \texttt{42026-F} and   \texttt{G32736-E}, and UNAM project No.  \texttt{IN113305}.
  %%%%%%%%%%%%%%%%%%%%%%%%%%%%%%%

%%% Appendix %%%  
\appendix
\section*{Appendix: Dark and Hall  conductivities.}  
\label{apen}

   In section \ref{kubo}  the method  to obtain the final expression for the  microwave induced conductance  \Eq{condLw}  was explained in detail. 
   Working along a similar procedure the expression for the  remaining conductivities are worked  from  
    equations (\ref{condd})  and (\ref{condw}).
   First we quote the  longitudinal dark conductance 
    %%%%%%%%%%%%%%%%%%%%%%%%%%
      \begin{equation} 
 \mathbf{\sigma}^D_{xx}   =  \frac{ e^2 \omega_c^2}{\pi \hbar}   \sum_\mu \, \mu \int d{\cal E}  \,  Im \, G_\mu\left( {\cal E}   \right)
 \,   \frac{d f }{d {\cal E}  } \,  Im \, G_\mu \left( {\cal E}  + \omega_c \right) \, , 
 \end{equation}
 %%%%%%%%%%%%%%%%%%%%%%%%%%
 whereas the dark  Hall conductance  is given by 
\hskip-2.0cm    \begin{equation}
 \mathbf{\sigma}^D_{xy}   =  \frac{ e^2 \omega_c^2 }{\pi \hbar  } \sum_\mu  \, \mu \int d{\cal E}  \, Im \, G_\mu\left( {\cal E}   \right)
 \left[ f\left( {\cal E} _\mu - \omega_c \right) -   f\left( {\cal E}  \right)\right]  \, P\, \frac{1}{\left({\cal E}  - {\cal E} _\mu + \omega_c \right)^2}
 \, ,   
 \end{equation}
 %%%%%%%%%%%%%%%%%%%%%%%%%%
 where $P$ indicates the principal-value integral. 
   The  final result for the microwave assisted 
 longitudinal conductivity was quoted in \Eq{condLw}. Following a similar procedure the microwave assisted 
 Hal conductivity is calculated to give 
 %%%%%%%%%%%%%%%%%%%%%%%%%% 
    %%%%%%%%%%%%%%%%%%%%%%%%%%
 \begin{equation}\label{condHw}
\hskip-2.5cm   \mathbf{\sigma}_{xy} ^\omega   =   \frac{ e^2 \omega_c^2 }{\pi \hbar }  
\int d{\cal E}     \sum_{\mu \nu}  \sum_l \sum_{mn}   Im \, G_\mu \left({\cal E}   \right)  
 \left[ f\left( {\cal E} _\nu  \right) -   f\left( {\cal E}  \right)\right] 
 T_{mn} \, \, \bigg{\vert} {\tilde J}_l\left( \vert \Delta_{mn} \vert \right)  V_{mn}  D_{\mu\nu} (\tilde{q}_{mn}) \bigg{\vert}^2     ,
 \end{equation}
 were the function $T_{mn}$ is defined as
\begin{equation} \label{funaux2} 
T_{mn} = \omega_c^3 \,    \, \frac{  (q_m^{(x)})^2 \, + \,   (q_n^{(y)})^2 }
   { \left(  {\cal E}  + \omega l  - {\cal E} _\nu    \right) \,\, {\vert \left({\cal E}  + \omega l -  {\cal E} _\nu    \right)^2 - \omega_c^2 \vert^2}} \, .   
 \end{equation}
  %%%%%%%%%%%%%%%%%%%%%%%%%%

\newpage

\end{document}